\documentclass{aa}
\usepackage[dvips]{graphicx}
\usepackage{txfonts}
\usepackage{longtable,lscape}
\usepackage{natbib}
\usepackage{cases}
\usepackage{subfig}
\usepackage[]{units}
\usepackage{fixltx2e}
\bibpunct{(}{)}{;}{a}{}{,}
\usepackage{color}
\usepackage{url}
\def\be{\begin{equation}}
\def\ee{\end{equation}}

\begin{document}

   \title{
First detection of the field star overdensity in the Perseus arm
}
 \author{M. Mongui\'{o} \inst{\ref{inst1},\ref{inst2}}
      \and P. Grosb\o{}l \inst{\ref{inst3}}
      \and F. Figueras \inst{\ref{inst2}}
          }
   \offprints{M. Mongui\'{o},
   \email{maria.monguio@ua.es}}
   \institute{ Departamento de F\'{\i}sica, Ingenier\'{\i}a de Sistemas y Teor\'{\i}a de la Se\~{n}al. Escuela Polit\'{e}cnica Superior, University of Alicante,
Apdo. 99, 03080 Alicante, Spain \label{inst1} \and
Departament d'Astronomia i Meteorologia and IEEC-ICC-UB,
     Universitat de Barcelona,
     Mart\'i i Franqu\`es, 1, E-08028 Barcelona, Spain \label{inst2} \and 
 European Southern Observatory, Karl-Schwarzschild-Str. 2, D-85748 Garching,
Germany \label{inst3} }
   \date{Received  / Accepted }
   \abstract
{}
{The main goal of this study is to detect the stellar overdensity associated with the Perseus arm in the anticenter direction.}
{We used the physical parameters derived from Str\"omgren photometric data to compute the surface density 
distribution as a function of galactocentric distance for different samples of intermediate young stars. The radial distribution of the interstellar absorption
has also been derived.}
{We detected the Perseus arm stellar overdensity at 1.6$\pm$0.2\,kpc from the Sun with a significance of 4-5$\sigma$ and a surface density amplitude 
of around 10\%, slightly depending on the sample used. Values for the radial scale length of the Galactic disk 
have been simultaneously fitted obtaining values in the range [2.9,3.5]\,kpc for the population of the B4-A1 stars.
Moreover, the interstellar visual absorption distribution is congruent with a dust layer in front of the Perseus arm.
} 
{This is the first time that the presence of the Perseus arm stellar overdensity has been detected through individual 
star counts, and its location matches a variation in the dust distribution.
The offset between the dust lane and the 
overdensity indicates that the Perseus arm is placed inside the co-rotation radius of the Milky Way spiral pattern.}

\maketitle

\section{Introduction}\label{intro}

The spiral arm structure in the Galactic disk is an important component when studying the morphology and dynamics of the Milky Way.
However, we still lack a complete picture  that describes the nature, origin, and evolution of this structure. 
Are there two \citep{2000A&A...358L..13D} or four \citep{2003A&A...397..133R} spiral arms? Are their main components gas or stars? 
Maybe, as was recently suggested by \cite{2008ASPC..387..375B},
there are two major spiral arms (Scutum-Centaurus and Perseus) with higher stellar densities and two minor arms (Sagittarius and Norma) mainly filled with gas and 
star forming regions. 
It is also important to understand how the stars interact with the arms: do they move through the arms \citep[density wave mechanism, ][]{1964ApJ...140..646L}, 
along the arms \citep[manifolds, ][]{2007A&A...472...63R}, or with the arms \citep[material arms, ][]{2012MNRAS.421.1529G}?

The Perseus spiral arm, with its outer structure placed near the Sun, is an excellent platform from which to undertake a study of its nature.
It has been studied considering different tracers such as HI neutral gas \citep{1967IAUS...31..143L}; large-scale CO distribution 
surveys \citep{2001ApJ...547..792D};  
a compilation of open clusters and associations and CO surveys of molecular clouds \citep{2008ApJ...672..930V};
star forming complexes \citep{2003A&A...397..133R,2014arXiv1405.7003F}; 
or masers associated with young, high-mass stars \citep{2006Sci...311...54X,2014ApJ...783..130R}.
Some of these studies trace the arm only in the second quadrant, and others trace it in the third quadrant, 
but few analyses  link both. Furthermore, few of them have enough data pointing towards $\ell=180\degr$, that is towards the anticenter.
Recently, \cite{2014ApJ...783..130R} published a substantial compilation of over 100 high-precision trigonometric distances to several masers using VLBI observations, in other words  tracing the location of star forming regions. Around 20 masers are located in the Perseus spiral arm, with five of them very close to the anticenter.
Their model for the spiral arms locates the Perseus arm at 2\,kpc at around $\ell=180\degr$,
with the five masers placed at slightly smaller galactocentric radius than the fit.
\cite{2014ApJS..215....1V} published a master catalog of the observed tangents to the Galaxy's  
spiral arms in order to fit a four-arm model. Using different arm tracers, he located the \textquotedblleft Perseus origin arm\textquotedblright  \,near the Galactic center at 
$\ell=338\degr$. The fitted model allows us to extrapolate the position of the Perseus arm at about 2\,kpc in the anticenter direction. 
Very interesting is his attempt to quantify the offset between different tracers, which favors the interpretation of the data in terms of the density wave theory.

The present study aims to trace the Perseus arm by studying both the radial stellar density variation toward the anticenter  with 
simple star counts, and by deriving the distribution of the radial velocity components through this direction.
Both arm structure and kinematics are essential issues in our study.
To trace the arm we use intermediate young stars with effective temperatures in the range [15000,9000]\,K (B4-A1 spectral type). 
These stars are excellent tracers of this overdensity as 1) they are bright enough to reach large
distances from the Sun, and 2) they are old enough to have had time
to respond to the spiral arm potential perturbation. 
 We explicitly avoid very young O-B3 and Cepheids since we aim to determine the location of the mass
peak (i.e., the potential minimum) associated with the Perseus arm and not the peak of star formation. In the density wave
scenario, one would expect a sequence (in distance or angle) starting with dust and then star formation, but offset with respect
to the potential \citep{1969ApJ...158..123R}. This scenario is in agreement with the recent work of \cite{2014ApJS..215....1V} where the hot dust
 (with masers and newborn stars) seems to peak near the inner arm edge while the stars are all over the arms.
Furthermore, to analyze the interaction between the arm and the stellar component,  the selected B4-A1 stellar tracers are 
young enough so their intrinsic velocity is still small, 
so that their response to a perturbation would be stronger and therefore easier to detect.
Our tracers --late B- or early A-type stars-- are  expected to show a 
density variation due to the presence of a perturbation. 
As mentioned, the most suitable direction to undertake this study on Galactic structure and kinematics is 
towards the anticenter  and the Perseus arm. First, this direction presents lower interstellar extinction than the direction pointing to the Galactic center.
Second, although slightly depending on the pitch angle,  this direction is the one expected to have the smallest distance to the 
Perseus arm stellar component. In addition, and more importantly, this direction 
was selected because the Galactic rotation would be negligible in the stellar radial velocity component spectroscopically derived for this young population, and it would directly 
reflect the velocity perturbation introduced by the spiral arm.

In the present paper we present the analysis of the stellar surface density through accurate distances and interstellar absorption derived 
from Str\"omgren photometry.  In a second paper of this series, work on the spiral arm kinematic perturbation 
will be provided utilizing
stellar radial velocity for a subsample of selected young stars.

In Sect. \ref{SAC} we overview the main characteristics of our photometric survey published and characterized in \cite{2013A&A...549A..78M,2014A&A...568A.119M}.
In Sect. \ref{WSsect} we carefully select the distance limited samples required for this study. 
All observational biases and constraints are evaluated and accounted for.
This analysis allows us to 
 estimate in Sect. \ref{arm} the local peaks in the surface density distribution compared to a pure exponential profile.
In Sect. \ref{OV} we characterize the stellar overdensity associated with Perseus, its distance from the Sun, the spiral arm amplitude, and its 
significance.
The differential interstellar absorption distribution and its relation with the Perseus spiral arm 
are discussed in Sect. \ref{layer}. Finally, in Sect. \ref{concl}, we summarize the main results and conclusions.

\section{Data}\label{SAC}

As published in \cite{2013A&A...549A..78M}, 
the authors carried out a $uvbyH\beta$ Str\"omgren photometric survey covering 16$\degr^2$ in the anticenter direction using the Wide Field Camera at the Isaac 
Newton Telescope (INT, La Palma).
This is the natural photometric system for identifying young stars and obtaining accurate estimates of individual distances and ages. 
The survey was centered slightly below the plane in order to take into account the warp, and covers Galactic longitudes from $\ell\sim$177$\degr$ to $\ell\sim183\degr$
and Galactic latitudes from $b\sim-2\degr$ to $b\sim1\fdg5$. 
The calibration to the standard system was undertaken using open clusters. 
We created a main catalog of 35974 stars with all Str\"omgren indexes and a more extended one with 96980 stars with partial data.  
The inner 8$\degr^2$ reach $\sim$90\% completeness at $V\sim17^{\unit{m}}$, while the outer sky area of $\sim$8$\degr^2$,
mostly observed with only one pointing, reaches this completeness at $V\sim 15\fm5$. Photometric internal precisions around 0.01-0.02$^{\unit{m}}$ 
for stars brighter than $ V=16^{\unit{m}}$ were obtained, increasing to 0$\fm$05 for some indexes and fainter stars ($V=$18-19$^{\unit{m}}$). 

In \cite{2014A&A...568A.119M} we describe in detail two different approaches implemented to
 compute the stellar physical parameters (SPP) for these stars.
The first  uses 
available pre-Hipparcos photometric calibrations  from the 1980s (empirical calibration method, EC) based on previous cluster data and trigonometric parallaxes. 
 This procedure follows two steps: 1) the classification of the stars in different photometric regions using extinction-free indexes ($[c_1], [m_1], H\beta, [u-b]$), and 2) for each region, 
the interpolation in empirical calibration sequences to obtain the intrinsic indexes and the absolute magnitudes from which interstellar extinction and distances can be computed.
The second method, which we developed, is based on the most accurate atmospheric 
models and evolutionary tracks available at present (model based method, MB). It starts with a 3D fit using three extinction free indexes  ($[c_1], [m_1], H\beta$), 
and their photometric errors. 
Then, the interpolation in evolutionary tracks provides the remaining physical parameters. Both methods were optimized for hot stars using the comparison with Hipparcos data.
In both cases, individual errors on SPP were computed through Monte Carlo random realizations.

The MB method provided distance, $M_V$, $A_V$, $(b-y)_0$, $T_{eff}$, and $\log g$, among other SPP for stars with $T_{eff}>$7000\,K.
On the other hand, the EC method provided distance, $M_V$, $A_V$, and $(b-y)_0$ for stars with spectral types in the range B0-A9.
By comparing these two sets of SPP data, a clear discrepancy of about $\sim$20\% in distance was detected and discussed.
The catalog, published in \cite{2014A&A...568A.119M}, also provided quality flags, some of 
 them related to the photometric classification required by the classical EC method \citep{1966ARA&A...4..433S} to disentangle
between early- and late-type stars. It was evident that the separation in the $[c_1]-[m_1]$ space is well defined and works properly
for stars with low photometric errors, but the gap between regions 
blurs when faint stars with larger photometric errors are considered. 
The IPHAS and 2MASS data, when available, were used to flag emission line stars and to check the coherence of some physical parameters derived using 
visual and infrared data, respectively. 
The fraction of binary stars in our sample may be important, in particular for hot stars. 
Thus in \cite{2014A&A...568A.119M} the effects of the secondary on the final accuracy of the photometric indexes and distance 
were also discussed.

\section{Distance complete samples}\label{WSsect}

\begin{figure}\centering
 \resizebox{\hsize}{!}{\includegraphics{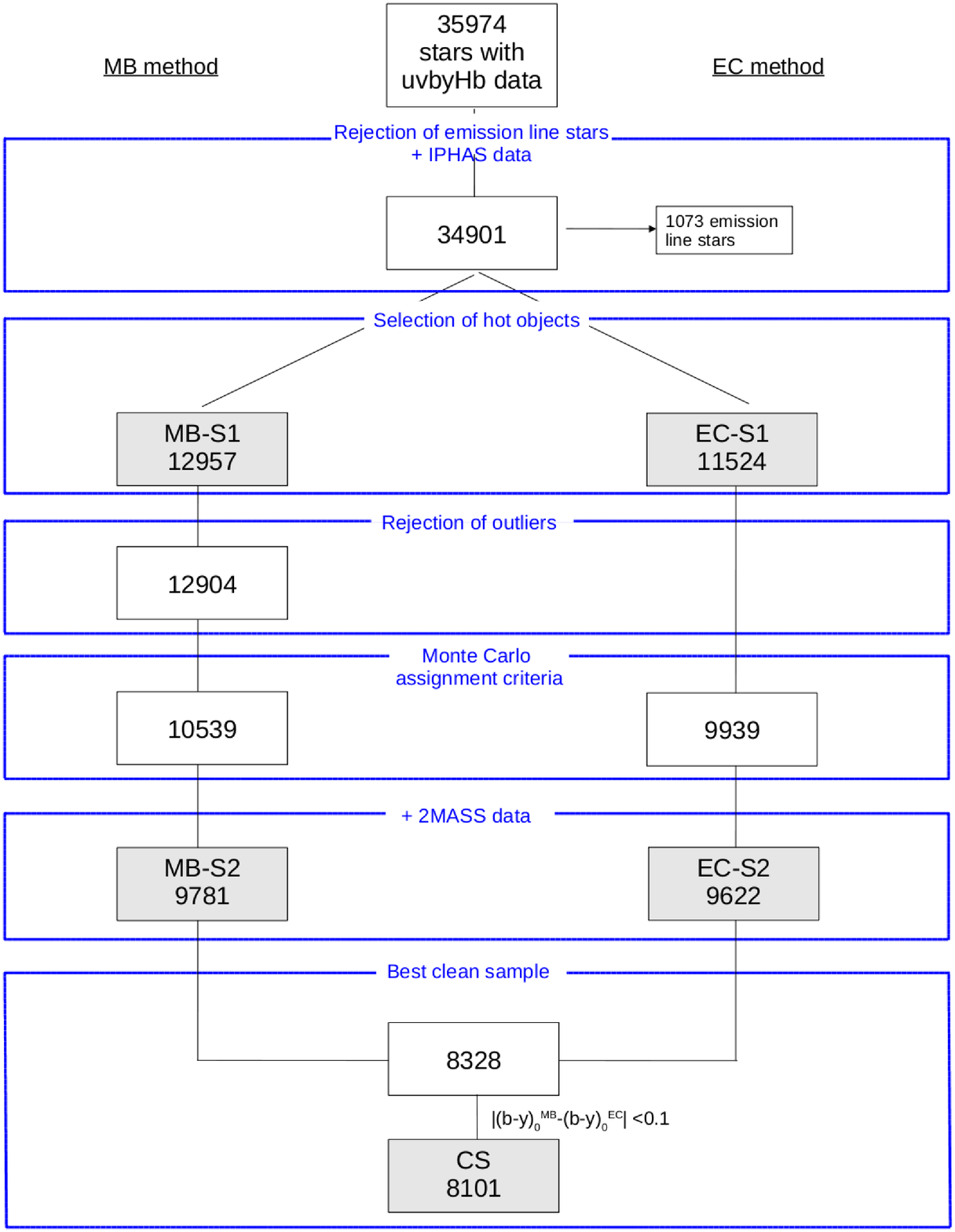}}
 \caption{Procedure for the generation of the working samples.
From top to bottom, the number of stars in the samples decreases while the quality of the individual physical parameters increases.
 The left side shows the samples according to the MB method, and the right side  with EC.}\label{WS}
\end{figure}

The available data --full $uvbyH\beta$ photometry for 35974 stars in the anticenter direction-- were cleaned and organized in 
different working samples by carefully evaluating both the 
quality of the data and the apparent magnitude and distance completeness (see  Fig.\,\ref{WS}).
In a first step, the available IPHAS initial data release (IDR) \citep{2008MNRAS.388...89G}, combined with our $H\beta$ index,
 allowed us to detect and reject 1073 emission line stars. 
Then each approach to compute the physical parameters (i.e., MB and EC) required its own strategy.
We created two different working samples: the first, hereafter 
MB-S1, with the 12957 stars with $T_{eff}>7000$K and SPP from the MB method,
and the second, hereafter EC-S1, with 11524  B0-A9 stars, with SPP derived using the EC method.
Following the MB strategy, 
 we considered as outliers the stars that lie 5$\sigma$ outside
the MB atmospheric grid during the 3D fit, 
taking into account the photometric errors in $[c_1]-[m_1]-H\beta$. 
We also removed stars with doubtful classification between photometric regions using $N_{side}$ 
and  $N_{reg}$ flags. 
These flags, $N_{reg}$ for the EC  method and $N_{side}$ for the MB method,
indicate the probability for a star to belong to earlier or later photometric regions and were computed through Monte Carlo simulations \citep[see][for details]{2014A&A...568A.119M}.
Coherence with 2MASS data allowed us to reject some stars that cannot be allocated reliably to either early (B0-B9) or late (A3-A9) types.
The two resulting  \textquotedblleft clean\textquotedblright \, samples contain 9781 (for MB-S2) and 
9622 stars (for EC-S2). A new subset was built containing the stars that simultaneously belong to the young population 
following both methods.  This was done by checking the coherence between the two sets of physical parameters. 
This cleanest sample contains 8328 stars and is named CS-MB or CS-EC depending on whether their physical parameters
were computed using the MB or EC method  (see Fig.\,\ref{WS}).
This sample contains stars with more accurate SPP data, but has fewer statistics owing to the
lack of stars rejected during the cleaning process.
From now on we will distinguish between the subsamples containing the stars in the inner sky area ($in$ suffix), 
from those where the stars located in the outer sky area are added ($all$ suffix).
Whereas the first subsample reaches fainter limiting magnitude, the second contains more stars but  is complete up to a brighter limiting magnitude.
 In Table \ref{Mvlim} we give the number of stars for each of the subsamples.

\begin{figure}\centering
 \resizebox{\hsize}{!}{\includegraphics{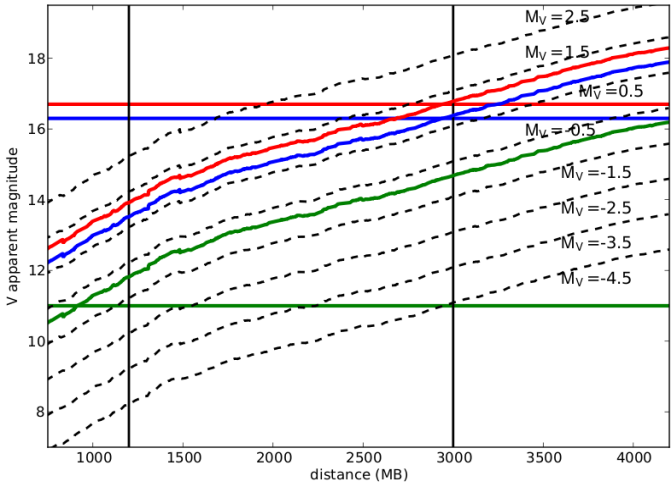}}
 \caption{Observed apparent magnitude for different absolute magnitude stars located at different distances. We have assumed an absorption of 
$\overline{A_V(r)} + \sigma _{A_V(r)}$ obtained from our data using the MB method. Vertical lines show the established limits at 1.2 and 3\,kpc.
Horizontal lines show the $V_{lim}$ computed for different working samples. In red, $V_{lim}^{max}$ and $M_{V lim}^{max}$ for MB-S1 and MB-S2 samples.
In blue, $V_{lim}^{max}$ and $M_{V lim}^{max}$ for CS-MB. In green, $V_{lim}^{min}$ and $M_{V lim}^{min}$  computed using $\overline{A_V(r)}$.}\label{VD}
\end{figure}

Since we wanted to derive how the stellar density varies with the galactocentric distance, we needed to ensure that the features observed are due 
to a physical and real overdensity and not to observational or selection biases. 
To account for this, we created distance complete samples by selecting ranges of absolute magnitude,
that ensured us distance completeness up to a given heliocentric distance. For each sample we estimated 
1) the maximum visual limiting magnitude $V_{lim}$, 
2) the magnitude at which some stars were saturated $V_{sat}$, and 
3) the maximum and minimum visual interstellar absorption at a given distance.
The $V_{lim}$ for each of the samples, assuming a 90\%
completeness, was estimated by computing the mean of the magnitudes at the peak star counts
in a magnitude histogram and its two adjacent bins, before and after the peak, weighted by the number of stars in each bin. 
In \cite{2013A&A...549A..78M} we checked that this estimate is adequate by comparing the $V$ magnitude distribution of the stars with all the
available photometric indexes, and with the distribution 
observed for stars measured in $V$ that reached a significant fainter limiting magnitude.
To ensure an adequate range before and after the Perseus arm in the anticenter direction, we selected samples complete 
up to 3\,kpc from the Sun.
The maximum interstellar absorption at 3\,kpc was estimated from $A_{Vmax}$(3\,kpc)=$\overline{A_V} + \sigma _{A_V}$. 
This value was then used to compute 
the limit in the intrinsic star brightness: $M_{V lim}(3\,\unit{kpc})=V_{lim}-A_{Vmax}(3kpc)-5\log(3kpc)+5$.
In addition, stars brighter than $V_{sat}=11^{\unit{m}}$ may suffer from saturation problems \citep{2013A&A...549A..78M}. 
Thus, we estimated and set a completeness limit in the absolute brightness --named $M_{V lim}(1.2 kpc)$-- for the closest distance, 
imposed to be 1.2\,kpc, 
using both $V_{sat}$ and the
interstellar absorption at this distance.
Following this strategy, that is, by selecting the stars with the absolute magnitude inside
the computed range,  all the resulting samples were converted to distance complete samples in the distance range 
between 1.2 and 3\,kpc.
Table \ref{Mvlim} shows all the values used to generate each sample together with the number of stars included in each of them.
Figure \ref{VD} shows an example of the apparent visual magnitude range needed to cover the distance range [1.2,3]\,kpc 
for a star with a fixed absolute magnitude.
As an example, this figure shows the limits on apparent visual magnitude assumed for the  MB-S1, MB-S2, and CS-MB samples. 
We can see how these samples have complete populations inside the heliocentric distance range selected.

Finally, we checked that the distribution of effective temperature and spectral type included in each of our samples
do not change as a function of distance within the [1.2,3]\,kpc range.  
Several two-sample Kolmogorov-Smirnov (KS) tests were done comparing the luminosity function at different distances with the 
distribution at the solar neighborhood (obtained in Sect. \ref{HMZP}, see Fig.\,\ref{SDZP}).
In all the cases, the KS tests are consistent with the assumption that the underlying luminosity functions arise from the same distribution
(see Table \ref{KS}). 
We observed that only for the farthest subsamples of MB-S1 and MB-S2  we do have small values of the p-values, but not enough 
to reject the null hypothesis at 0.01 level.

Looking at Table \ref{Mvlim} we see that the samples considering $all$ the anticenter observed sky area, 
as the outer area has a brighter limiting magnitude, have a very narrow range in absolute magnitude, which means that   the final number of stars in those samples is not high enough to undertake the study of the stellar overdensity 
induced by the Perseus arm with good statistical significance (Sect. \ref{arm}).

\begin{table*}
\caption{$A_{V}$, $V_{lim}$, and $M_{Vlim}$ for all samples at the bright (min) and faint (max) ends. 
Spectral type according to the $M_{Vlim}$ are shown. N is the number of stars for the 
 absolute magnitude limited samples, i.e., $M_{Vlim}($1.2\,kpc$)<M_V<M_{Vlim}($3\,kpc$)$, also indicating  which of them are between 1.2 and 3\,kpc.
N/N$_{S1}$ is the ratio between the number stars (between 1.2 and 3\,kpc) of each sample and the S1 sample, always using the $M_V$ range of the sample.
}\label{Mvlim}
\centering\begin{tabular}{c|c|c|c|c|c|c|c|c|c|c|c|c}
 & \multicolumn{6}{c|}{MB}&\multicolumn{6}{c}{EC}\\\hline
 & \multicolumn{3}{c|}{in}&\multicolumn{3}{c|}{all}&\multicolumn{3}{c|}{in}&\multicolumn{3}{c}{all}\\\hline
& S1 & S2 & CS & S1& S2 & CS&S1 & S2 & CS & S1& S2 & CS\\\hline\hline
$V_{lim}^{min}$&\multicolumn{12}{c}{11}\\\hline
$V_{lim}^{max}$&16.7&16.7&16.3&15.7&15.7&15.3&16.7&16.7&16.3&15.7&15.7&15.3\\\hline\hline
$A_{Vmin}$(1.2kpc)&\multicolumn{6}{c|}{1.5}&\multicolumn{6}{c}{1.8}\\\hline
$A_{Vmax}$(3kpc)&\multicolumn{6}{c|}{3.1}&\multicolumn{6}{c}{3.2}\\\hline\hline
$M_{Vlim}$(1.2kpc)&\multicolumn{6}{c|}{-0.9}&\multicolumn{6}{c}{-1.2}\\\hline
SP$_{min}$& \multicolumn{6}{c|}{B4}&\multicolumn{6}{c}{B3.5}\\\hline\hline
$M_{Vlim}$(3kpc)&1.2 & 1.2 & 0.8 & 0.2& 0.2 & -0.2&1.1 & 1.1 & 0.7 & 0.1& 0.1 & -0.3\\\hline
SP$_{max}$&A1 & A1 & A0 & B8.5& B8.5 & B7.5&A0 & A0 & B9.5 & B8.5& B8.5 & B7\\\hline
$N_{tot}$&7763&6061&4998&12957&9781&8101&6756&5943&4998&11524&9622&8101\\\hline
N($M_{Vlim})$&2209 & 1565 & 672 & 1587& 935 & 278&821 & 653 & 361 & 446& 289 & 133\\\hline
N (1.2-3kpc)&851 & 733 & 349 & 403& 326 & 106 & 441 & 378 & 185 & 151 & 111 & 33\\\hline
N/N$_{S1}$  &1.00&0.86 &  0.65 &1.00&0.81 & 0.63 & 1.00&0.86 & 0.82 & 1.00& 0.74& 0.57\\
\end{tabular}
\end{table*}

\begin{table*}
\caption{Two sample KS test results of the comparison between solar luminosity function and luminosity function at different distances ranges.
We provide the KS statistic $D_{n_{1},n_{2}}$, the number of stars for each of the samples ($n_1$ for the solar sample, $n_2$ 
for the sample at each distance), and the p-value.}\label{KS}
\centering\begin{tabular}{c|c|ccc|ccc|ccc|ccc}
       &&\multicolumn{3}{c|}{1.2-1.5\,kpc}&\multicolumn{3}{c|}{1.5-2.0\,kpc}&\multicolumn{3}{c|}{2.0-2.5\,kpc}&\multicolumn{3}{c}{2.5-3.0kpc}\\
       &$n_1$&          $D_{n_{1},n_{2}}$&$n_2$         &  p     &      $D_{n_{1},n_{2}}$&$n_2$    &    p          &    $D_{n_{1},n_{2}}$&$n_2$         &        p           &   $D_{n_{1},n_{2}}$&$n_2$         &  p     \\\hline
MB-S1in& 343 &      0.10&91& 0.44 & 0.09&231&0.20 &0.06&236&0.73& 0.12&293& 0.03\\
MB-S2in& 343 &      0.11&81& 0.44 & 0.09&200&0.27 &0.08&202&0.45& 0.12&250& 0.03\\
CS-MBin& 195 &      0.16&40& 0.37 & 0.09&89 &0.75 &0.15&98 &0.09& 0.12&122& 0.23\\
EC-S1in&  36 &      0.15&62& 0.61 & 0.16&129&0.43 &0.10&134&0.90& 0.16&116& 0.46\\
EC-S2in&  36 &      0.21&55& 0.27 & 0.16&116&0.47 &0.13&117&0.69& 0.14& 90& 0.67\\
CS-ECin&  18 &      0.24&20& 0.56 & 0.25& 62&0.31 &0.14& 56&0.94& 0.13& 47& 0.98\\  
          \end{tabular}
\end{table*}

\section{Radial density distribution}\label{arm}

The surface density was chosen over volume density
to be the best parameter with which to trace the stellar distribution considering the information
 available, that is, a set of distance complete samples inside a solid angle of 8$\degr^2$ or 16$\degr^2$ in the anticenter direction 
(Sect. \ref{WSsect}). 
This  is a better choice  since it gives an estimate of the total disk mass at a given distance, also taking into account the 
change in the mid-plane due to the warp, assuming that the disk scale height varies slowly with distance.
All the factors taken into account in the derivation of this parameter from the data are described in 
Sect. \ref{SDcomp}. The radial bin size used to derive the galactocentric variation of this parameter was computed using the method
proposed by \cite{2006physics...5197K}. As stated by the author, this method is optimized to find substructure in the 
data\footnote{Nonetheless, we checked that a  simpler procedure such as the constant radial bin of 200\,pc also provides  congruent
results with equivalent conclusions.}. 
The radial scale length of the Galactic disk in the anticenter direction and the possible overdensity associated with Perseus were 
obtained by fitting an exponential function to the working samples described in Table \ref{Mvlim}. 
Initially, the two parameters of this function were fitted: the radial scale length ($h_R$) and the surface density in the solar neighborhood ($\Sigma_{\odot}$) of the 
population represented by these working samples. The fit highly depends on the zero point of the distribution $\Sigma_{\odot}$. 
Thus, to substantially increase both the range in distance covered and the significance of the fit, 
we decided to derive  this parameter independently using a local sample (see Sect. \ref{HMZP}). Then these values were used in Sect. \ref{RSL} 
to do the fits, where the departure of the exponential surface density distribution and the existence of possible maxima are discussed in terms of 
the location and existence of the overdensity associated with the Perseus spiral  arm. 

\subsection{The computation of the stellar surface density}\label{SDcomp}
To derive the stellar surface density we included the sky area surveyed, the effect of the scale height of the underlying thin 
disk population being considered, as well as the presence of the Galactic warp.
The surface density for each distance bin was computed as 
\begin{center}
 \begin{equation}\label{sig2}
\Sigma (r_k) = \frac{1}{S_k}\sum_{i=1}^{n_k}\frac{1}{F_{Z,i}},
\end{equation}\end{center}
where $n_{k}$ is the number of stars in a given radial bin $k$;  $S_k$ is the disk projected surface for that radial bin 
computed from the Galactic latitude $\ell$ covered, 
the mean distance to the Sun $r_k$ of the bin, and the distance width size of the bin in parsecs; 
$F_{Z,i}$ is a correction factor for each star that takes into account the fact that our observed solid angle does not cover the full
vertical cylinder. This factor is assumed to depend on the vertical density distribution, modeled  as  
$sech^2(z/h_z)$ \citep{1981A&A....95..105V}, with $h_z$ being the scale height of the disk. 
As is known, $h_z$ depends on the age of the population. 
Here, as an approximation, we model it as depending on the visual absolute magnitude of the star.
Values for $h_z$ are not well determined,
whereas \cite{2000AJ....120..314R} gave $h_z=$25-65\,pc for OB-type stars (mainly O-B2), 
and \cite{2001AJ....121.2737M} obtained $h_z$=34.2$\pm$2.5\,pc
for a sample of O-B5 stars. More recently, \cite{2008ChA&A..32..360K}, using Hipparcos data, estimated values around
$h_z$=103.1$\pm$3.0\,pc for the full range of OB stars. 
\citet{2014A&A...564A.102C} adopted  $h_z$=130 pc for stars with ages younger than 0.15\,Gyr whereas values up to 260\,pc were assumed for stars with ages between 0.15 and 1\,Gyr.
These simulations provide age ranges around 50$\pm$20\,Myr  for B5-type stars ($T_{eff}\sim$ 15000\,K),  
while age ranges of 500$\pm$200\,Myr are obtained for A5-type stars ($T_{eff}\sim$ 8000\,K). 
Taking into account these estimations, we adopted the relation
$h_z(\unit{pc})=36.8\cdot M_V+130.9$, which is equivalent to assuming 100\,pc for B5-type stars and 200\,pc 
for an A5-type star. It must be taken into account that a
different relation between the scale height and the intrinsic 
brightness would change the zero point of the stellar density distribution, and thus the corresponding surface density at the Sun's 
position $\Sigma_{\odot}$. 
The $F_{Z,i}$ factor also takes into account the vertical range $(z_{min,i},z_{max,i})$  covered at each 
galactocentric distance. Keeping this in mind, this factor is computed following
\begin{center}
\begin{equation}\label{EqFz}
 F_{Z,i}(r)=\frac{\int_{z_{min,i}}^{z_{max,i}} sech^2 \left(\frac{z-z_W(r)}{h_z(M_{V,i})}\right) dz}{\int_{-\infty}^{+\infty} sech^2\left(\frac{z-z_W(r)}{h_z(M_{V,i})}\right)dz},
\end{equation}
\end{center}
where we also considered the position of the warp ($z_W(r)$).
From observational data (2MASS and HI data) \cite{2006A&A...451..515M} obtained $b_W\sim$-0.5$\degr$ at $\ell=180\degr$,
so we can assume that at different distances the $z_W$, where the star density of the disk is maximum, can be computed as 
$z_W=r \cdot \tan b_W$.

\subsection{Surface density in the solar neighborhood}\label{HMZP}
\begin{figure}
\resizebox{\hsize}{!}{\includegraphics{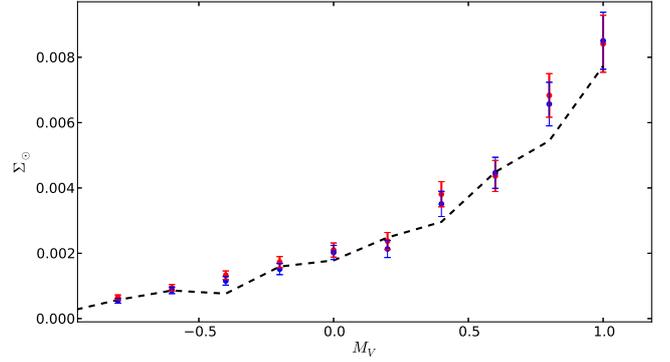}}
 \caption{Surface density at the Sun's position for 0.2$^{\unit{m}}$ $M_V$ bins, using both $V_{lim}=6^{\unit{m}}$ (in red) and $V_{lim}=6\fm4$ (in blue). 
In black we have the same values as obtained from the Besan\c{c}on Galaxy model.
}\label{SDZP}
\end{figure}

We  used the \cite{1998A&AS..129..431H} catalog of Str\"omgren photometry to compute the surface density at the Sun's 
position for young stars. This process was done
 following the same methodology
as we used for the anticenter stars, that is, the same computation of SPP, and the same process used to clean the samples 
(Sects. \ref{SAC} and \ref{WSsect}).  
First we checked the completeness of the \cite{1998A&AS..129..431H} catalog  in terms of the visual apparent magnitude.
For that, we cross-matched this catalog with the 
Hipparcos catalog (ESA 1997). 
We verified that the completeness is above 95\% up to 6$\fm$5 for OB stars. 
Str\"omgren photometric data allowed us to compute their physical 
parameters, selecting only those with $T_{eff}>7000\unit{K}$ for MB, and B0-A9 for EC. 
To mimic the samples in the anticenter we selected the same $M_V$ ranges, 
and computed the distance limit for which we can ensure completeness (assuming in this case $A_V=0$): 
$r_{lim}=10^{(V_{lim}-M_{Vmax}+5)/5}$. We repeated the computations using $V_{lim}=6^{\unit{m}}$
and $V_{lim}=$6$\fm$4 and obtained very similar results. 
We considered the effect of the warp to be negligible in the solar neighborhood.  
 We also checked that  different values for the distance of the Sun above 
the Galactic plane gave  results   within the error bars: $z_{\odot}=$15\,pc from \cite{2003A&A...409..523R} and $z_{\odot}$=26\,pc from \cite{2009MNRAS.398..263M}. 
We used the same dependence of the scale height function on absolute magnitude as for the 
density in the anticenter direction (Sect. \ref{SDcomp}). In this case, the limits in $z$ follow the surface of an sphere around the Sun's position, 
 $z_{max}=-z_{min}=\sqrt{r_{lim}^2-x^2-y^2}$, and the projected surface density is then easily modeled as 
$S_k=\pi r_{lim}^2$. 
The obtained distribution for different magnitudes ranges of 0$\fm$2
is plotted in Fig.\,\ref{SDZP}. 
For comparison, the values resulting from model B of the new Besan\c{c}on Galaxy Model 
\citep{2014A&A...564A.102C} are overplotted. These values were derived fitting the model to the full sky Tycho catalogue. 
In Fig.\,\ref{SDZP} we observe 
that the distribution is very similar in  the whole range of absolute magnitudes, thus confirming the robustness of the derivation
of the local surface density for the young population, i.e., the zero point of our exponential fit to the anticenter data. 
When we used the $M_V$ limits established for our samples (see Table \ref{Mvlim}), 
the values for the local surface density obtained were (3.25$\pm$0.13)$\cdot10^{-2}$ for MB-S1in and MB-S2in, 
(1.70$\pm$0.07)$\cdot10^{-2}$ for CS-MBin, (3.49$\pm$0.12)$\cdot10^{-2}$ for EC-S1in and EC-S2in, 
and (1.41$\pm$0.06)$\cdot10^{-2}\,\star$/pc$^2$ for CS-ECin.
Their errors were computed as
\begin{center}
 \begin{equation}
 \sigma_{\Sigma}=\frac{\sqrt{n}}{\Sigma_{i=1}^n F_{Z,i} S_k/n}.
\end{equation}
\end{center}

\subsection{The radial scale length and the detection of the Perseus arm}\label{RSL}
\begin{figure*}
\resizebox{\hsize}{!}{\includegraphics{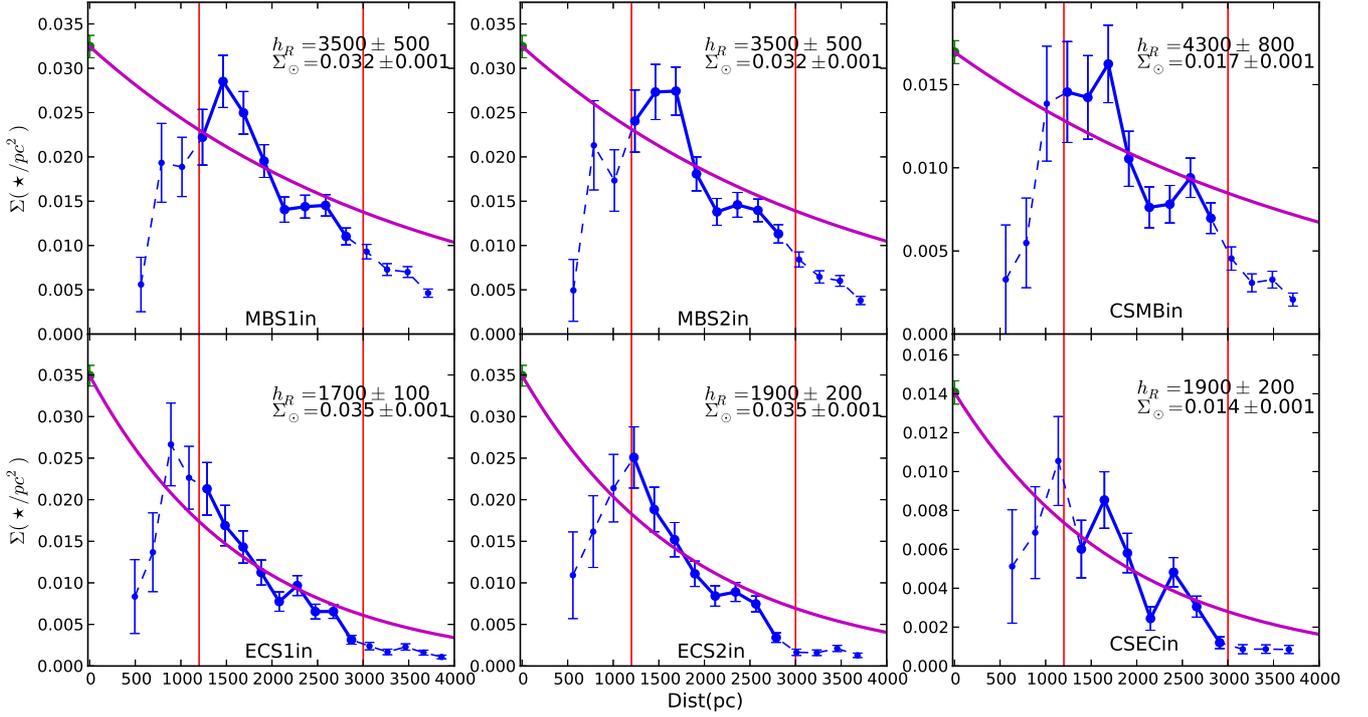}}
 \caption{Radial variation of the stellar surface density for the MB (top) and EC (bottom) samples in blue.  Vertical red lines show the 1.2 and 3\,kpc completeness limits.
The exponential fit is plotted in magenta, with the $h_R$ and $\Sigma_{\odot}$ parameters expressed 
in pc and $\star$/pc$^2$, respectively. Blue dots joined with solid lines are the ones used for the fit, since it is the region where completeness is ensured.
}\label{FSD}
\end{figure*}

The radial surface density distribution was computed for all the samples 
and an exponential function was fitted to each of them by adopting the zero point values for the local surface density obtained in 
the previous section.
The radial scale length of the Galactic disk in the anticenter direction is presented in Fig.\,\ref{FSD}.
The scale lengths obtained for the fits using MB distances (Fig.\,\ref{FSD}, top panel)  are in the range 
$h_R\sim$[3.5-4.3]\,kpc.
More importantly, we can clearly recognize the overdensity associated with the Perseus arm at around 1.6\,kpc (depending slightly on the sample), 
within the distance range at which the samples are complete, that is, between 1.2 and 3\,kpc from the Sun.  
The decline after 3\,kpc is produced by the limiting magnitude of our samples, since we carefully checked that the completeness
is only ensured up to this distance. 
The results for the EC method are presented in the bottom panels of Fig.\,\ref{FSD}. 
As discussed in Sect. \ref{SAC}, the photometric distances derived from this method are clearly biased,
in the sense that EC distances can be 20\% shorter than the MB distances, the latter being more congruent with Hipparcos parallaxes 
\citep[the detailed comparison is given in][]{2014A&A...568A.119M}. 
As a consequence of this bias, the overdensity bump associated with the Perseus arm 
is also present, but closer to us and overlapping with the short distance limit at 1.2\,kpc imposed to avoid saturation effects from bright stars. 
Some attempts were made to fit the points beyond the arm, but the results obtained for the radial scale length are unrealistic
($h_R$=[1.7,1.9]\,kpc). 
For the samples using $all$ the sky area, the number of stars in the range [1.2-3]\,kpc was too low, 
since the limiting magnitude was too bright, and the 
fainter stars included in the samples have $M_{V}$=0.2,-0.2, i.e., B7-B8.5 stars (see Table \ref{Mvlim}). 
Thus, the features related to the overdensity, although present, were less statistically 
significant than for the $in$ samples.

\begin{figure*}
\resizebox{\hsize}{!}{\includegraphics{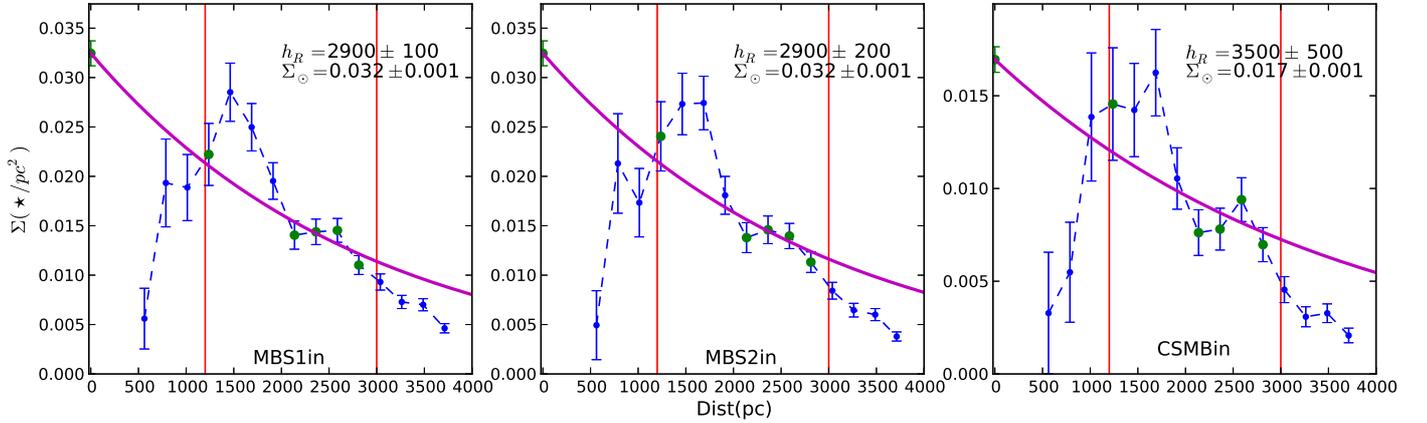}}
 \caption{Radial variation of the stellar surface density for the MB samples in blue.  Vertical red lines show the 1.2 and 3\,kpc completeness limits.
Green dots show the points used for the exponential fit, avoiding those around the overdensity location.
The exponential fit is plotted in magenta, with the $h_R$ and $\Sigma_{\odot}$ parameters expressed 
in pc and $\star$/pc$^2$, respectively.
}\label{FSDna}
\end{figure*}

The next step was to re-compute the exponential fitting, avoiding the points close to the peak detected, 
that is between 1.4 and 2.0\,kpc (see Fig.\,\ref{FSDna}).
In this fit, a slight decrease of the radial scale length until $h_R=$2.9$\pm$0.1/0.2\,kpc for both MB-S1 and MB-S2 samples is obtained. 
For the cleanest sample CS-MB we obtain a slightly larger value of $h_R=$3.5$\pm$0.5\,kpc (right panel of Fig.\,\ref{FSDna}).
In this case, the uncertainty is significantly larger since the working sample contains fewer stars.
In addition, the stars in this sample have a brighter apparent limiting magnitude with respect to the previous ones, 
so their corresponding
 $M_V$ range is shifted to intrinsically brighter objects, i.e., it contains a slightly younger population on average.
This result is congruent with recent data obtained from simulations such as those presented by \citet{2013ApJ...773...43B}, 
showing that the radial scale length increases when the population considered is younger.

The three samples using MB distances show a clear peak around 1.6\,kpc (Figs. \ref{FSD} and \ref{FSDna}).
This overdensity is  clearer for those  samples having larger number of stars (MB-S1 and MB-S2, left and central panels). 
Although they have a more robust statistical detection, they also contain values that are less accurate for the SPP and some contamination 
from adjacent photometric regions could also be present (see Sect. \ref{SAC}).
Thus we consider that it is the whole set of fits presented in the top rows of Figs. \ref{FSD} and \ref{FSDna}, 
using the three samples, 
that allows us to confirm the detection of the stellar overdensity associated with the Perseus spiral arm.

\section{The Perseus arm overdensity}\label{OV}
\begin{figure}\centering
\resizebox{0.9\hsize}{!}{\includegraphics{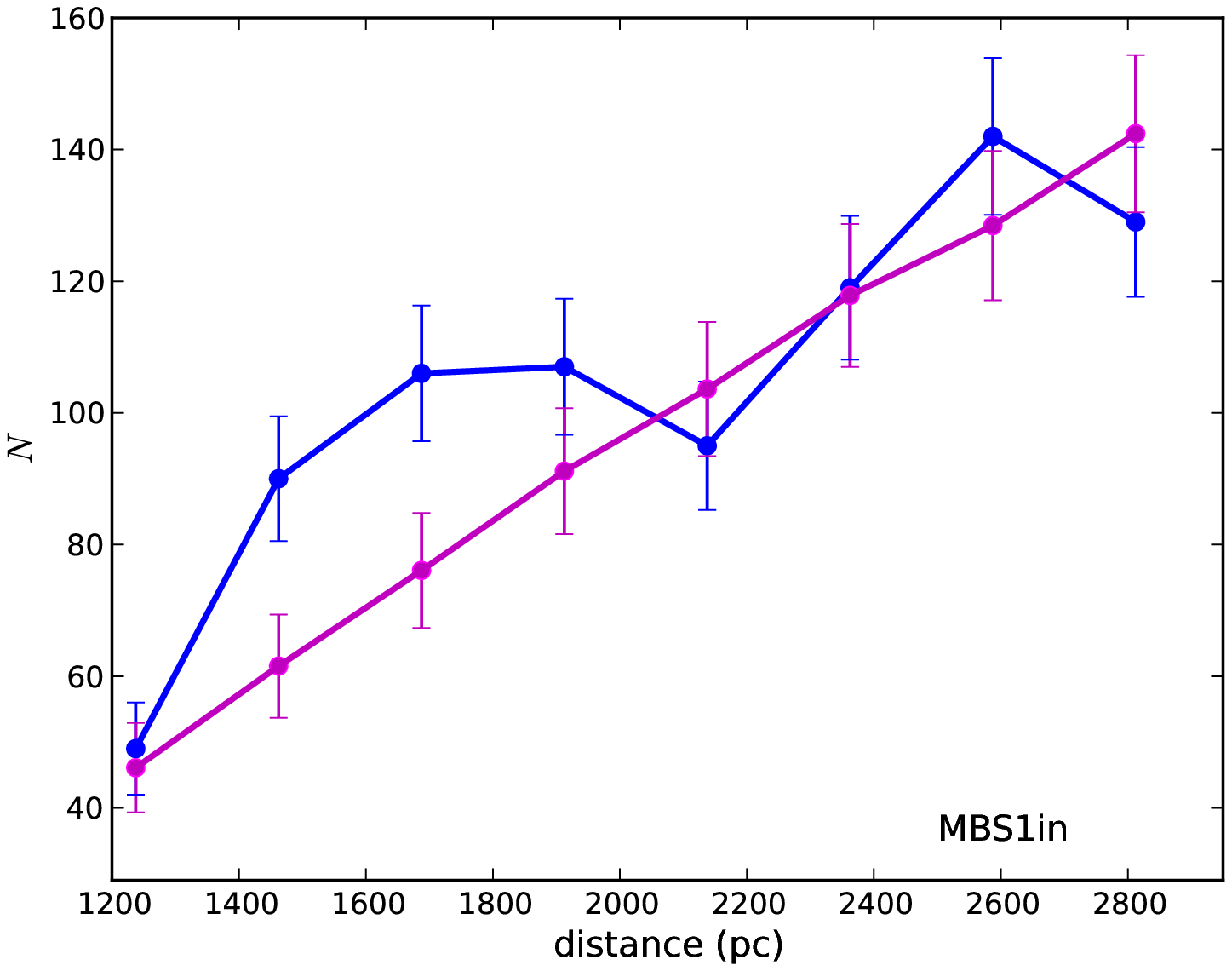}}
 \caption{Number of stars $n_k$ observed (in blue) and number of stars expected from the exponential fit 
$n_{k}^{fit}$ developed in Fig.\,\ref{FSDna} (in magenta) as a function of distance for the MBS1in sample.
}\label{SDtoN}
\end{figure}

We performed $\chi^2$ tests for the different fits presented in the previous sections to quantitatively evaluate the significance of the 
detected Perseus arm overdensity.
The number of stars observed ($n_{k}^{obs}$, with $k=1,m$) for each $k$ distance bin between 1.2 and 3\,kpc was compared with the 
distribution coming from the resulting exponential fit (see Fig.\,\ref{SDtoN}).
To compare the model and the observations we use as a first approximation
the number of corresponding stars at the central distance of each bin ($n_{k}^{fit}$).
The $\chi^2$ value was computed as
\begin{equation}
\chi^2=\sum_{k=1}^m \frac{\left(n_{k}^{obs}-n_{k}^{fit}\right)^2}{n_{k}^{fit}},
\end{equation}
where $n_{k}^{fit}$ was computed from the surface density obtained from the fitted expression $\Sigma_{k}^{fit}$ and transformed to 
the number of stars per bin 
using: $n_{k}^{fit}=\Sigma_{k}^{fit}\cdot S_k \cdot\langle F_{Z,i}\rangle$. The factor $\langle F_{Z,i}\rangle$ is the average of 
$F_{Z,i}$ for all the stars in the bin previously used
 to transform from observed $n_k$ to observed surface density $\Sigma(r_k)$ (see Eq. \ref{sig2}). 
Then, considering the degrees of freedom for each histogram (i.e., number of bins minus two, since $h_R$ and $\chi^2$ have been 
estimated), 
the test statistics were obtained.
The hypothesis that the data come from a pure exponential distribution
can be rejected at a 4-5$\sigma$ confidence level 
for S1, S2, and CS samples for the inner sky area, 
so they clearly do not fit with an exponential. Looking at the plots, the overdensity at 1.6\,kpc 
is the main reason for the deviation. The location of the peak of the distribution varies slightly  for different samples, being at 
1.5\,kpc for MB-S1in, and 1.7\,kpc for MB-S2in and CS-MBin.
An error of 0.2\,kpc (the bin width) has been adopted for this maximum overdensity.

On the other hand, when we rejected the points between 1.4 and 2.0\,kpc  (where the overdensity is detected) 
and repeat the fit, we obtained 
p-values from the $\chi^2$ test of 0.44, 0.50, and 0.10 for the S1, S2, and CS samples (those in Fig.\,\ref{FSDna}). 
In other words, when 
we did not take into account the bins located close to the arm the distributions were fully compatible 
with an exponential fit.

We estimated the significance of the peak as $(n_{arm}^{obs}-n_{arm}^{fit})/\sqrt{n_{arm}^{obs}}$, where $n_{arm}^{obs}$ is the number of stars in the bins close to the
arm (i.e., between 1.4 and 2.0\,kpc) and $n_{arm}^{fit}$ is the expected number of stars from the 
exponential fit (where the arm bins were not used for the fit).
The overdensity obtained for the MB-S1 and MB-S2 samples had a significance of 4.3$\sigma$. For CS-MB we obtained 3.0$\sigma$, a lower value
due to the smaller number of stars included in the sample (see Table \ref{TSig}).

The amplitude of the arm $A$ was estimated from the expression  $A=(n_{arm}^{obs}-n_{arm}^{fit})/(n_{arm}^{obs}+n_{arm}^{fit})$. 
The obtained values are presented in Table \ref{TSig}. For all samples we derive values in the range $A$=[0.12,0.14],
well in agreement with recent determinations \citep[see references of the density contrast of the Milky Way spiral arms
discussed in][]{2011MNRAS.418.1423A}. 
However, one should take into account the different methods used by these authors: 
\cite{2001ApJ...556..181D} obtained $A$=0.14 
from K band surface brightness, while \cite{2005ApJ...630L.149B} derived  $A$=0.13 from GLIMPSE data looking at 
the tangential points of the inner spiral arm.
 In external galaxies, \cite{1995ApJ...447...82R} gave values in the range
  0.15$\lesssim A\lesssim$0.6, while \cite{2004AA...423..849G} provided amplitudes for several galaxies, also from
 surface photometry in the $K$ band, reaching values up to $A=$0.5, 
although they claimed that the amplitude varies as a function of radius.
However, the same authors  \citep{2014ASPC..480..117G}   suggested that amplitudes measured in the K band may be 
overestimated by a factor of 2 which bring 
the Perseus amplitude in agreement with measurements of external galaxies.

\begin{table}
 \caption{Number of stars  between 1.4 and 2.0\,kpc, and the expected value from the exponential when we do not take into account the arm bins.
Significance and amplitude of the arm are also indicated for each of the three MB samples.}\label{TSig}
\centering\begin{tabular}{c|c|c|c|c}
&$n_{arm}^{obs}$&$n_{arm}^{fit}$&Signif.&$A$\\\hline
MB-S1in&303&228&4.3&0.14 \\
MB-S2in&302&227&4.3&0.14\\
CS-MBin&182&142&3.0&0.12\\
 \end{tabular}
 \end{table}

\section{The Perseus arm dust layer}\label{layer}
Since both individual photometric distances and visual interstellar absorptions ($A_V$) are available for a 
large sample of stars, a detailed 3D extinction map was created 
to discuss and quantify the possible existence of a dust layer related to the Perseus arm. In Sect \ref{ExtMap} we 
present the maps and discuss some of the observed features, whereas in Sect.\,\ref{layerb} the existence of a change in the differential absorption ($dA_V/dr$)
is discussed in order to determine if there is a high density interstellar region 
linked to the Perseus arm and, more importantly, to understand if this region is in front of or behind the stellar component
of this arm.

\subsection{Three-dimensional extinction map in the anticenter direction}\label{ExtMap}

\begin{figure*}\centering
 \resizebox{0.8\hsize}{!}{\includegraphics{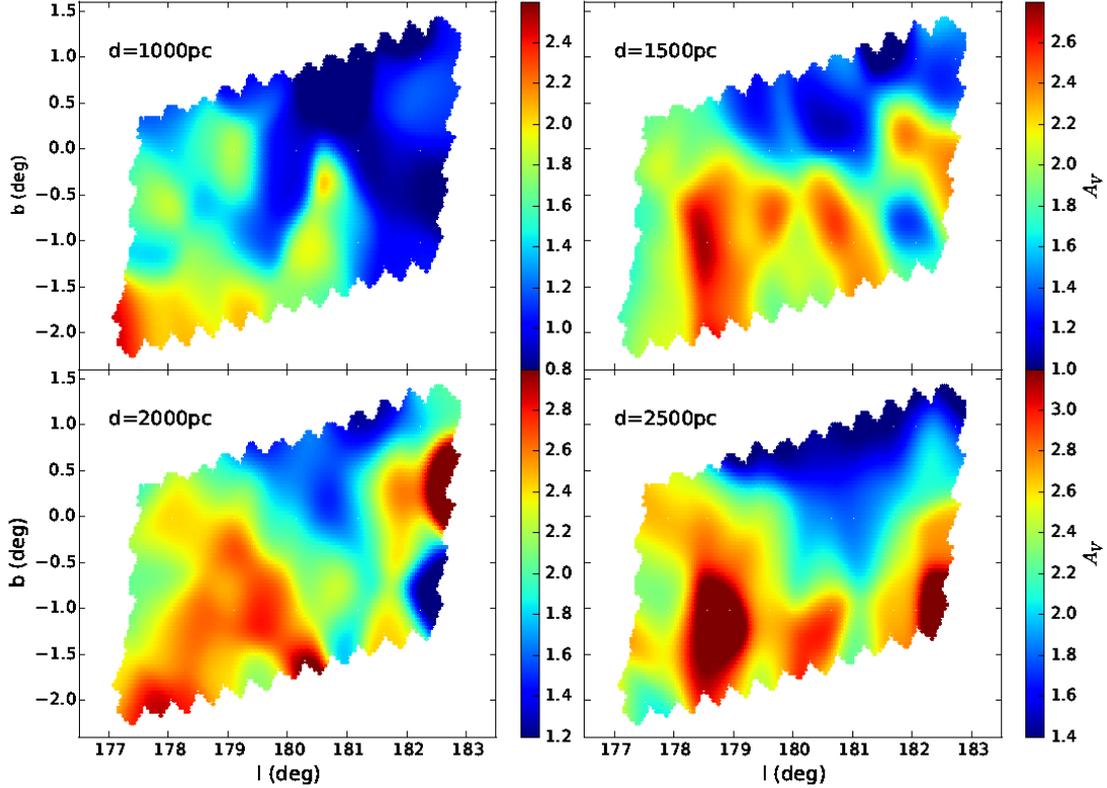}}
 \caption{2D extinction maps in $(\ell,b)$ at four different distances from the Sun, 1000, 1500, 2000, and 2500. 
$A_V$ is color coded, with a shift of 0.2$^{\unit{m}}$ in the color scale between consecutive plots. 
}\label{Av3D1}
\end{figure*}

The MB-S1 sample was used to create the 3D extinction map presented in Fig.\,\ref{Av3D1}.
The number of stars in this sample allowed us to reach distances up to 2.5-3\,kpc.
The grid inside the 3D cone was constructed taking 20\,pc steps in distance and 2.5\,arcmin steps in both
 galactic longitude and latitude. At each point of the grid, the absorption $A_V$ was 
computed as
\begin{equation}\label{EqAvm}\centering
 A_{V_{(r,l,b)}}=\frac{ \sum_{i=1}^{N} {A_{V_i} \exp{\left(\frac{ -\Delta r_i ^2 }{2\sigma^2}\right)}}}{\sum_{i=1}^{N} \exp{\left(\frac{ -\Delta r_i ^2 }{2\sigma^2}\right)}}
,\end{equation}
where $N$ is the total number of stars in the sample and $\Delta r_i$ is the distance between the $ith$ star and the point of the grid.
The $\sigma$ value is the radius used for the Gaussian weight computed as $\sigma=r\cdot atan (\alpha)$ with 
$\alpha=0.25\degr$. 
This method allowed us to obtain a grid with higher spatial resolution in nearby regions where we had more 
information, and less at further distances where information was poorer.
 We note that this method would introduce some bias at the edges of the grid.
The lack of data outside the surveyed area can result in mean values of the $A_{V_{(r,l,b)}}$ that are slightly biased towards
the values present in the inner parts.
We estimate that this bias should be present only in areas at angular distances less than about 0.5$\degr$ (2$\sigma$) from the survey's  edges.

\begin{figure}\centering
 \resizebox{\hsize}{!}{\includegraphics{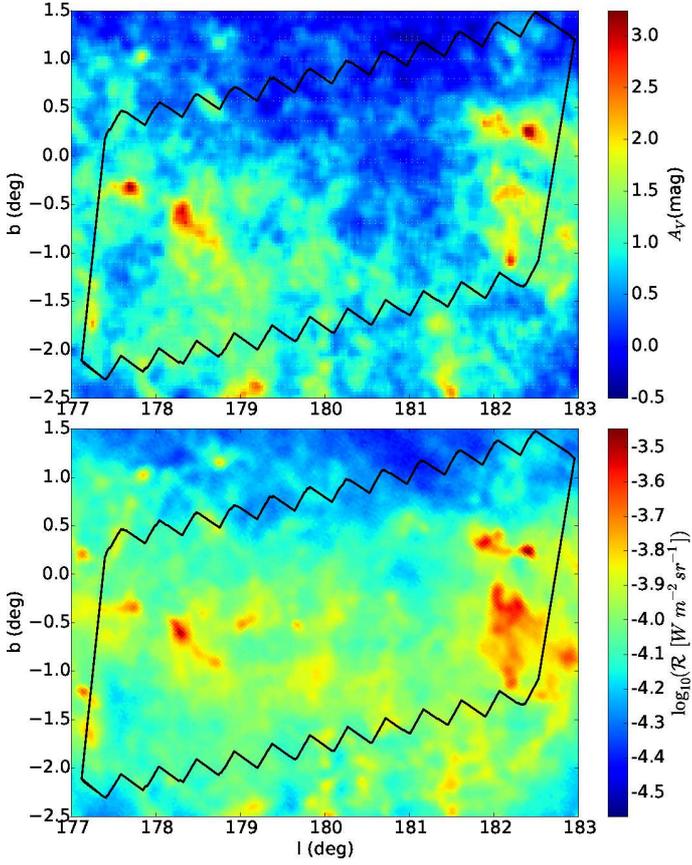}}
 \caption{Top: Extinction map from \cite{2007MNRAS.378.1447F} based on 2MASS data. 
Bottom: thermal the dust radiance $\mathcal{R}$ map (or dust integrated intensity)
from Planck data \citep{2014A&A...571A..11P}. 
}\label{Planck}
\end{figure}

In Fig.\,\ref{Av3D1} we plotted the $A_V$ values of the grid points placed at four distances and computed  following 
Eq.\,\ref{EqAvm}. 
A first comparison can be done with the integrated extinction maps obtained by \cite{2007MNRAS.378.1447F}, using 
2MASS data (Fig.\,\ref{Planck}, top), and those recently derived by the Planck collaboration (2014; Fig.\,\ref{Planck}, bottom).
As expected the same general trends are observed in these maps clearly identifying the same very low and very high extinction 
areas. Whereas both maps in Fig.\,\ref{Planck} indicate the total dust integrated intensity towards a given line of sight,
our 3D  map is able to distinguish features at specific distances. Although our data avoids the presence of some of the biases thoroughly 
discussed in \cite{2007MNRAS.378.1447F}, we note that it is not the goal of this paper to investigate   
the individual distribution of clouds or their sizes and mass distribution, which are  discussed in that paper.
In the Froebrich maps, for example,  regions close to rich star clusters would be dominated by the colours of the cluster 
members;  instead, the data presented here do not suffer from this bias. 
As an example of the future scientific exploitation of this 3D map we here comment on the distribution of interstellar medium around  the supernova remnant (SNR)  Simieis 147.
This SNR has its geometrical center at 
(l,b)=(180.17, b=-1.81) with a diameter of about 180\,arcmin \citep{2009yCat.7253....0G}. 
Its suggested distance ranges from 0.6 to 1.9\,kpc \citep[see Table 1 from][]{2015arXiv150107220D}. 
For its related pulsar PSRJ0538+2817 \cite{2007ApJ...654..487N} estimated a distance of $1.47^{+0.42}_{-0.27}$\,kpc and 
\cite{1985ApJ...292...29F} an absorption of $A_V=0.76\pm 0.2$ magnitudes.
At the edge of our survey our 3D extinction map indicates 
a slightly higher absorption at larger distances (at about\,2 kpc) and a more prominent absorption at the area surrounding the geometrical center proposed by \cite{2009yCat.7253....0G}.
As a second example we discuss the features visible around $(\ell,b)\sim(182.0,0.0)$ in the 1.5\,kpc  and  2\,kpc maps.
\cite{2006ApJ...652.1230R} found a molecular cloud in the anticenter direction associated with the two IRAS sources  IRAS 05431+2629
at $(\ell,b)=(182.1,-1.1)$ and IRAS 05490+2658 at $(\ell,b)=(182.4,+0.3)$.
This last HII region was also found by \cite{1982ApJS...49..183B} at $(\ell,b)=(182.36,+0.19)$ through CO observations, 
and they give a distance of 2.1$\pm$0.7\,kpc.
We observed a weak but well-defined high mean  density region at this distance.  
These are only examples of the potential interest of these maps, and it will be 
through future analysis that multivariate clustering techniques will be applied to our grid points to confirm and characterize 
the clumpy  structures  present in the cone defined by our anticenter survey.

Recently, \cite{2014MNRAS.443.1192C}  obtained a 3D extinction map with a spatial resolution of  3-9\,arcmin  
 by using multiband photometry. In this study  individual distances are not available, so  to break the degeneracy between 
the intrinsic stellar colours and the amounts of extinction the authors relied on the combination of  optical and near-infrared 
photometric data. Our maps presented in Fig.\,\ref{Av3D1}, although with less spatial resolution, are based on  individual distances 
with accuracies at about 10-20\% \citep{2014A&A...568A.119M}. We understand that both studies provide complementary information for future 
analysis. 

\begin{figure}\centering
 \resizebox{\hsize}{!}{\includegraphics{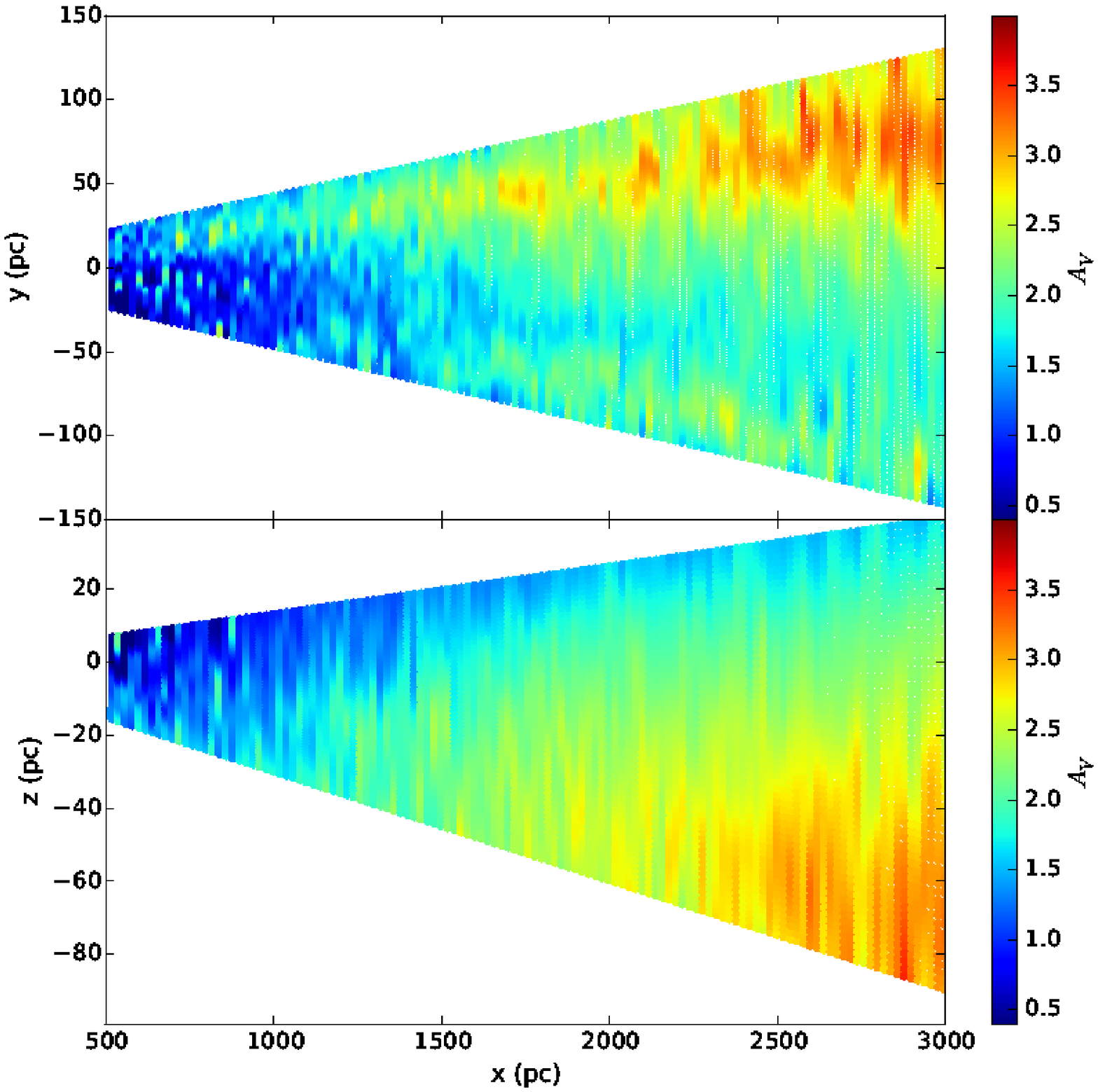}}
 \caption{Top: 2D extinction map in $(x,y)$, where stars for all $z$ were used for the average.
Bottom: 2D extinction map in $(x,z)$, where absorption for the stars at all $y$ were averaged.
$A_V$ is color coded. }\label{XYm}
\end{figure}

As a final example,
in Fig.\,\ref{XYm} we show the 2D extinction map projected on the cartesian coordinates $(x,y)$ and $(x,z)$ planes where the absorptions at different $z$ and $y$ were averaged. 
The $x$-axis is positive toward the anticenter, $y$ in the direction of galactic rotation, while $z$ toward the north galactic pole.
 These maps allowed us to analyze the change of extinction with distance. We observed that larger extinction 
regions were placed at lower galactic longitudes and below the Galactic plane.
This large-scale  dust distribution is discussed in the next section in the context of the presence of the Perseus spiral arm.

\subsection{Absorption distribution as a function of distance}\label{layerb}

\begin{figure*}\centering
 \resizebox{0.9\hsize}{!}{\includegraphics{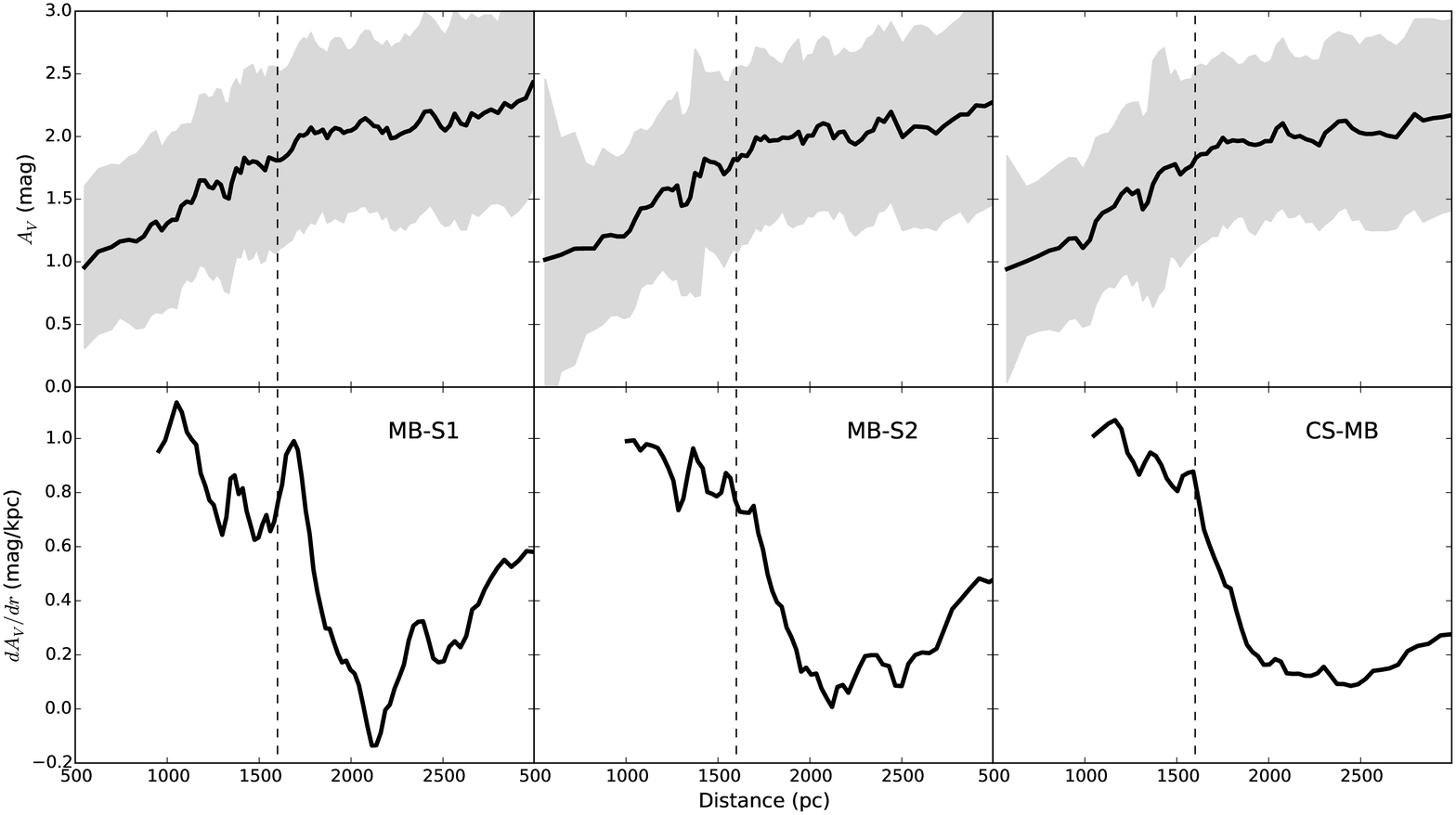}}
 \caption{Top: Running median of the visual absorption vs. heliocentric distance distribution obtained from the three different samples MB-S1 (left), MB-S2 (middle), and CS (right). 
The gray area indicates the error of the median computed as $1.253\,\sigma / \sqrt{N}$.
Bottom: differential visual absorption showing  the 
change in slope of the upper plots. Vertical dashed lines show the location of the Perseus arm at 1.6\,kpc derived in Sect.\,\ref{arm}.}
\label{Avdist}
\end{figure*}

Here, to reinforce the detection of the stellar overdensity of the Perseus arm presented in Sect.\,\ref{OV},
we present in Fig.\,\ref{Avdist} (top) the median  distribution 
of the visual absorption as a function of distance. We can clearly appreciate a change in the slope
at the same position where we detected the stellar overdensity of Perseus, which is at about 1.6\,kpc. 
This change is better identified by plotting its derivative as a function of distance ($dA_V/dr$) (see Fig.\,\ref{Avdist}, bottom).
We observed that this derivative presents two different well-defined values, a high value of about 0.8-1.0\,mag/kpc in front of the 
stellar arm, and a significantly lower value of about 0.2-0.4\,mag/kpc behind the arm. 
We would expect, for constant absorption, a flat distribution of the $dA_V/dr$, while
the presence of a cloud or dust layer would be translated into a bump in the $dA_V/dr$ distribution. 
We can see this change in slope in all the directions, suggesting the presence of a dust layer 
just before the position of the density peak associated with the Perseus arm.

As described by \cite{1969ApJ...158..123R}, the  star formation induced by the shock in a stellar density wave scenario 
would produce an azimuthal age sequence across 
the spiral arms. 
However, as a first attempt, we analyze the dust distribution (i.e., interstellar extinction) in this context.
The presence of the dust lane at the inner side of the stellar overdensity indicates a compression or shock in the gas 
associated with the spiral arm
\citep[although the strict coincidence of a shock and a lane is not obligatory;][]{2004MNRAS.349..909G}. 
From this we can use our first analysis of the 3D extinction map to establish a limit on the co-rotation (CR) radius in the Milky Way 
Galactic disk  \citep{1972ApJ...173..259R}. 
A discussion on the optical tracers of spiral wave resonances in external galaxies can be found in \cite{1992ApJS...79...37E}. 
\cite{1997ApJ...476L..73P} provided a good sketch of the Z-trailing spatial arms configuration expected for the Milky Way
 (see their Fig.\,1).  Figure \ref{Avdist} (right) suggests that the dust layer is just in front of the Perseus arm so 
the CR radius of the spiral pattern (if a density wave scenario is assumed)  is outside the location of Perseus in the anticenter direction: 
  $R_{CR}>$ 10.2\,kpc (assuming $R_{\odot}=8.5$\,kpc).  
This result is closely in agreement with the position of the CR radius suggested by \cite{2011MNRAS.418.1423A},
who, from a kinematic analysis of the moving groups in the solar neighborhood, proposed an angular velocity
 of the perturbation ($\Omega_p$) of about 16-20\,km$\cdot$s$^{-1}\cdot$kpc$^{-1}$, 
so with a galactocentric co-rotation radius at about $R_{CR}$=11-14\,kpc. 

\section{Summary and conclusions}\label{concl}
In \cite{2013A&A...549A..78M}  and \cite{2014A&A...568A.119M} we published a deep Str\"omgren photometric survey and a new strategy 
for the derivation of stellar physical parameters for the young stellar population in the anticenter direction. 
Here we used
these data to derive samples of intermediate young stars, complete within the distance range between 
1.2 and 3.0\,kpc from the Sun. 
We computed the surface density distribution in the anticenter 
direction by assuming a vertical density distribution of the disk, taking into account the surveyed area and the warp,
so carefully defining correction functions for all the observational and physical effects.
The same strategy was used to compute 
the surface density at the Sun's position through \cite{1998A&AS..129..431H} Str\"omgren data, obtaining
0.032$\pm$0.001$\star/\unit{pc}^2$ and 0.017$\pm$0.001$\star/\unit{pc}^2$ for stars in the $M_V=$[-0.9,1.2] and 
$M_V=$[-0.9,0.8] ranges, respectively.
The exponential functions fitted  to the computed stellar surface density
revealed a clear star overdensity at around 1.6$\pm$0.2\,kpc that we associate with the presence of the Perseus spiral arm. 
The obtained distance is slightly closer than the 2.0\,kpc recently obtained by 
\citet{2014ApJ...783..130R} who traced the arm using star forming regions.
The fits we performed avoiding the bins close to the arm location, allowed us to derive the radial scale length of the Galactic 
disk young population, obtaining $h_R=$2.9$\pm$0.1\,kpc for B4-A1 
stars, and $h_R=$3.5$\pm$0.5\,kpc for B4-A0 stars. 
The overdensity associated with the Perseus arm has been detected with a significance above 3$\sigma$ in all cases, 
and above 4$\sigma$ for the samples with larger statistics.
Our results indicate that the star density contrast of the young population in Perseus 
is on the order of $A$=0.12-0.14.
This quantitative estimation of the spiral arm amplitude in the anticenter --derived here for the first time--
  should be understood as a contribution to the future derivation of the amplitude changes of the Milky Way spiral pattern 
as a function of galactocentric radius. 
So far, this radial change has only been clearly detected in external galaxies \citep[e.g.,][]{2004AA...423..849G}.

Also very important, the distribution of visual interstellar absorption as a function of distance reveals the presence of a 
dust layer 
in front of the Perseus arm, which suggests that it is placed inside the co-rotation radius of the Milky Way spiral pattern.
Our results locate the co-rotation radius at least two kiloparsecs outside the Sun's position.
This is the first time that the presence of the Perseus arm has been detected through individual 
star counts, and its presence is being supported by the dust distribution.

\begin{acknowledgements}
This work was supported by the MINECO (Spanish Ministry of Economy) 
- FEDER through grant AYA2009-14648-C02-01 and CONSOLIDER CSD2007-00050.
M. Mongui\'{o} was supported by a Predoctoral fellowship from the Spanish Ministry (BES-2008-002471 through ESP2006-13855-C02-01 project).
\end{acknowledgements}

\bibliographystyle{aa}
\bibliography{MGF3-pb}

\end{document}